\documentclass[prl,aps,twocolumn,groupedaddress,floats,superscriptaddress,nopacks]{revtex4-2}
\usepackage{graphicx}
\usepackage{subfigure}
\usepackage{amssymb}
\usepackage{latexsym}
\usepackage{epsfig}
\usepackage{psfrag}
\usepackage{dcolumn}
\usepackage{amsmath,amsfonts,amssymb,bm,ulem}
\usepackage{color}
\usepackage{float}

\usepackage{soul}
\usepackage[dvipsnames]{xcolor}

\usepackage{changes}

\usepackage{changes}

\newcommand{\be}{\begin{equation}}
\newcommand{\ee}{\end{equation}}
\newcommand{\bea}{\begin{eqnarray}}
\newcommand{\eea}{\end{eqnarray}}

\newcommand{\rrs}[2]{{\color{Blue}\st{#1}{#2}}}

\begin{document}
\title{Interaction-enhanced nesting in Spin-Fermion and Fermi-Hubbard models}

\author{R. Rossi}
\affiliation{Sorbonne Universit\'{e}, CNRS, Laboratoire de Physique Th\'{e}orique de la Mati\`{e}re Condens\'{e}e, LPTMC, F-75005 Paris, France}
\affiliation{\'Ecole Polytechnique F\'ed\'erale de Lausanne (EPFL), Institute of Physics, CH-1015 Lausanne, Switzerland}
\author{F. \v{S}imkovic IV}
\email{current address: IQM GERMANY GmbH, 80636 M\"{u}nchen, Germany}
\affiliation{CPHT, CNRS, Ecole Polytechnique, IP Paris, F-91128 Palaiseau, France}
\affiliation{College de France, 11 place Marcelin Berthelot, 75005 Paris, France}
\author{M. Ferrero}
\affiliation{CPHT, CNRS, Ecole Polytechnique, IP Paris, F-91128 Palaiseau, France}
\affiliation{College de France, 11 place Marcelin Berthelot, 75005 Paris, France}
\author{A. Georges}
\affiliation{Center for Computational Quantum Physics, Flatiron Institute,
162 Fifth Avenue, New York, New York 10010, USA}
\affiliation{College de France, 11 place Marcelin Berthelot, 75005 Paris, France}
\affiliation{CPHT, CNRS, Ecole Polytechnique, IP Paris, F-91128 Palaiseau, France}
\affiliation{DQMP, Universit\'{e} de Gen\`{e}ve, 24 quai Ernest Ansermet, CH-1211 Gen\`{e}ve, Switzerland}
\author{A. M. Tsvelik}
\affiliation{Condensed Matter Physics and Materials Science Division, Brookhaven National Laboratory, Upton, NY 11973-5000, USA}
\author{N.V. Prokof'ev}
\affiliation{Department of Physics, University of Massachusetts, Amherst, MA 01003, USA}
\author{I.S. Tupitsyn}
\email{itupitsyn@physics.umass.edu}
\affiliation{Department of Physics, University of Massachusetts, Amherst, MA 01003, USA}


\begin{abstract}
The spin-fermion (SF) model postulates that the dominant coupling between low-energy fermions in near critical metals is mediated by collective spin fluctuations (paramagnons) peaked at the N\'{e}el wave vector, ${\bf Q}_N$, connecting hot spots on opposite sides of the Fermi surface. It has been argued that strong correlations at hot spots lead to a Fermi surface deformation (FSD) featuring flat regions and increased nesting. This conjecture was confirmed in the perturbative self-consistent calculations when the paramagnon propagator dependence on momentum deviation from ${\bf Q}_N$ is given by $\chi^{-1} \propto |\Delta q|$. Using diagrammatic Monte Carlo (diagMC) technique we show that such a dependence holds only at temperatures orders of magnitude smaller than any other energy scale in the problem, indicating that a different mechanism may be at play. Instead, we find that a $\chi^{-1} \propto |\Delta q|^{2}$ dependence yields a robust finite-$T$ scenario for achieving FSD. To link phenomenological and microscopic descriptions, we applied the connected determinant diagMC method to the $(t-t')$ Hubbard model and found that in this case: (i) the FSD is not very pronounced, and, instead, it is the lines of zeros of the renormalized dispersion relation that deform towards nesting; (ii) this phenomenon appears at large $U/t>5.5$ before the formation of electron and hole pockets; (iii) the static spin susceptibility is well described by $\chi^{-1} \propto |\Delta q|^{2}$. Flat FS regions yield a non-trivial scenario for realizing a non-Fermi liquid state.
\end{abstract}

\maketitle


\noindent \textbf{Introduction.} The two-dimensional spin-fermion model was formulated as an effective theory for near critical metals and used to explain the non-Fermi liquid physics and superconducting instability in doped cuprates \cite{Chub2000,Chub2003}. In this semi-phenomenological model the low-energy quasiparticles interact with collective magnetic fluctuations, or paramagnons. In some theories of high-$T_c$ superconductors the paramagnons play a role similar to phonons in ordinary metals, with the general concept formulated in Refs.~\cite{Hertz1976,Millis1993}. The model has found a wide range of applications in cuprates and iron-based superconductors including proposals for pairing mechanisms in the ``strange metal" \cite{Wang2017}.

On approach to the antiferromagnetic instability, the spectrum of paramagnons softens and strong near-critical fluctuations may cause the Fermi surface (FS) to change its shape near hot spots---these are FS points connected by the N\'{e}el wave vector ${\bf Q}_N$. In what follows we consider the case with eight hot spots when ${\bf Q}_N = (\pm \pi, \pm \pi)/a$ with $a=1$ being the lattice constant. The exchange interaction mediated by soft modes changes the dispersion relation and, thus, deforms the FS in the vicinity of hot spots making it flatter and closer to the nesting condition between the opposing FS patches (for illustration see, Fig.1 in \cite{Tsvelik-1-2017}). This remarkable Fermi surface deformation (FSD) is solely the effect of strong correlations within the hot spots. However, when is comes to establishing the quantitative description of the phenomenon, one finds that the result crucially depends on the form of the paramagnon propagator $\chi $. Its frequency dependence is dominated by Landau damping, $\propto |\omega|$, describing decay of spin waves into particle-hole excitations \cite{Chub2003,Millis1994}. The momentum dependence is more subtle, and different behaviors lead to different results.

The critical Ornstein-Zernike form, $\chi^{-1} \propto |\Delta {\bf q}|^2$, with $\Delta {\bf q} = {\bf q}-{\bf Q}_N$, leads to the non-Fermi liquid behavior \cite{Chub2000,Chub2003,Millis1994,MeSa2010,Sachdev2012,Maier2016,Gerlach2017}; the tendency to FSD was established in one-loop RG \cite{Chub2000,Chub2003}. [A model with four hot spots, when the FS touches the antiferromagnetic Brillouin zone boundary (AFBZB) is a ``non-conventional'' Fermi liquid state; its self-energy differs from the standard Fermi liquid expectations but the low-energy quasiparticles remain well defined \cite{Tremblay2012,Wang2013}.] It was suggested in Ref.~\cite{Lee2017} that in the limit of weak coupling and small angle between the ${\bf Q}_N$ and FS normal at hot spots, the effective paramagnon propagator takes the form $\chi^{-1} \sim |\Delta q|$. In this case, the outcome of the second-order self-consistent solution is a flow towards flat hot spot domains. However, an attempt of Ref.~\cite{Berg2020} to verify predictions of Ref.~\cite{Lee2017} using determinant Monte Carlo simulations \cite{determSC,Berg2012} at temperature $T/E_F \geq\ 0.005$ did not find evidence for linear momentum dependence $\chi^{-1} \sim |\Delta q|$ despite observing deviations from the Ornstein-Zernike ansatz.

It should be noted that since interactions mediated by critical modes do not displace quasiparticles along the FS, well nested flat regions effectively act as one-dimensional subsystems \cite{Tsvelik-1-2017}. This is an alternative scenario for realizing the non-Fermi liquid state that explains several important properties of the pseudogap regime in cuprates, including the excitation spectrum and transport phenomena (see, for instance, \cite{Tsvelik-2-2017,Tsvelik2019}).

In this Letter, we address the FSD problem within the spin-fermion and $(t-t')$ Hubbard models in the parameter regime considered optimal for high-$T_c$ superconductivity. Using diagrammatic Monte Carlo (diagMC) method we first verify the self-consistent scenario of Ref.~\cite{Lee2017} for the SF model and demonstrate that it holds only at ultra-low temperatures. Next, to connect phenomenological considerations with microscopic physics, we apply the connected determinant diagMC technique (CDet) to the $(t-t')$ Hubbard model to reveal how FSD develops in this case and what is the corresponding  momentum dependence of the spin susceptibility.

The diagMC approach combines advantages of quantum field-theory tools with power of MC sampling of complex configuration spaces~\cite{DiagMC1,DiagMC2}. It works directly in the thermodynamic limit and does not suffer from the conventional fermionic sign problem~\cite{Signproblem}, making it suitable for solving systems with arbitrary dispersion relations and shapes of the interaction potential~\cite{TMNP2016,Dirac2017,LeBlanc2022}. The CDet approach~\cite{Rossi2017} performs efficient summation of all diagram topologies ``on the fly" and currently represents the most advanced unbiased technique for the Hubbard model.

\smallskip

\noindent \textbf{Models.} In the SF model with hot spots, Landau damping renders the ``bare" quadratic dependence of $\chi$ on frequency irrelevant at low temperature. This allows one to formulate the model in the Hamiltonian form \cite{Chub2000,Chub2003}:
\begin{eqnarray}
H &=& \sum_{{\bf k}, \alpha} \epsilon_{\bf k} c^{\dag}_{{\bf k},\alpha} c^{\,}_{{\bf k},\alpha}
+ \sum_{q} \chi^{-1}_0(q) {\bf S}_{q} \rrs{}{\cdot}{\bf S}_{-q} \nonumber \\
&+& g \sum_{{\bf k}, {\bf q}, \alpha, \beta, i} c^{\dag}_{{\bf k+q},\alpha} \sigma^i_{\alpha, \beta} c^{\,}_{{\bf k},\beta} S^i_{-q},
\label{HSF}
\end{eqnarray}
where {$c^{\dag}_{\bf k,\alpha}$} is the fermion creation operator in the state with momentum ${\bf k}$ and spin projection $\alpha$, $\sigma^{i}$ are Pauli matrices ($i=x, \; y, \; z$), $g$ is the SF coupling constant, $\epsilon_{\bf k}$ is the bare electron dispersion, and ${\bf S}$ is the collective spin degree of freedom. In what follows we choose $\epsilon_{\bf k}$ as the tight-binding dispersion relation on the square lattice with two hopping amplitudes, $t$ and $t'$ ($t=1$ in our units), yielding $\epsilon_{\bf k} = -2 t (\cos k_x + \cos k_y) - 4 t' \cos k_x  \cos k_y$.

Feynman diagrams for proper self-energy, $\Sigma(\textbf{q},\omega_n)$, and polarization, $\Pi(\textbf{q},\omega_m)$, where $\omega_n$ and $\omega_m$ are fermionic and bosonic Matsubara frequencies, respectively, follow from the standard rules for the conventional and the skeleton representations, and have the same structure as for fermions interacting via
\begin{equation}
V_{i,j}({\bf q},\omega_m) = - \frac{g}{4}
\frac{ \sigma^z_i \sigma^z_j + (\sigma^{+}_i \sigma^{-}_j + \sigma^{-}_i \sigma^{+}_j)/2 }
{ \gamma |\omega_m| + \chi^{-1}_0(|{\bf q}-{\bf Q}_N|) },
\label{SFI}
\end{equation}
where $\gamma$ is the Landau damping constant (in what follows we take $\gamma/t=1$). This form respects the spin $SU(2)$-symmetry (on the square lattice all four vectors $(\pm \pi, \pm \pi)$ are formally identical to ${\bf Q_N}$).

The standard Fermi-Hubbard model on a square lattice at half filling is parameterized by the nearest- and next-nearest-neighbor hopping amplitudes $t$ and $t'$ and on-site repulsion $U$ (see, for instance, Ref.~\cite{SRF2022}). For description of the corresponding diagrammatic series and their treatment see Refs.~\cite{Rossi2017,SRF2022,Wietek2021}.

\smallskip

\begin{figure}[t]
\centerline{\includegraphics[scale=0.35]{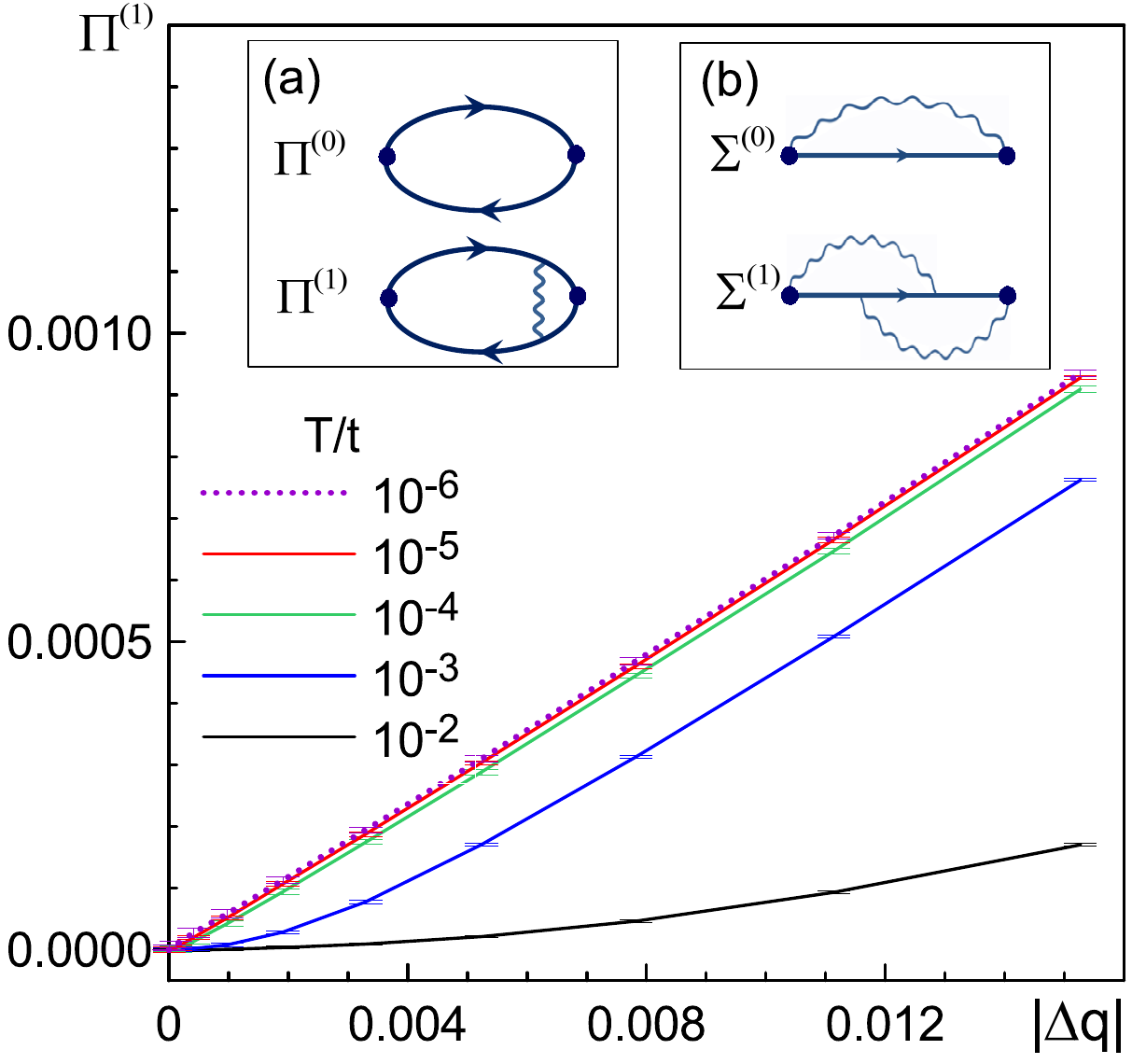}}
\caption{Polarization $\Pi^{(1)}$ as a function of $|\Delta {\bf q}|$ (see text) along the $(\pi, \pi)$
direction at zero Matsubara frequency. Insets show leading-order diagrams contributing to (a) polarization function and (b) proper self-energy. All data were obtained for $t'/t=-0.1$, $c=1$, $g=1$, and $\eta = 10^{-6}$ at half-filling; in this case the initial angle $\theta_N$ is less than $10^{\circ}$).}
\label{Fig1}
\end{figure}

\noindent \textbf{SF model.} We start by verifying the scenario put forward in Ref.~\cite{Lee2017}, which corresponds to
the limit of weak coupling and small angle $\theta_N$ between ${\bf Q}_N$ and the FS normal at the hot spot. Building on the one-loop RG result of Ref.~\cite{MeSa2010} that dynamical exponent $z$ decreases from $z=2$  it was conjectured that the flow is towards $\chi^{-1} = [ c(|\Delta {\bf q}_x| + |\Delta { \bf q}_y|) + |\omega| ]$. Moreover, it was shown that $\chi^{-1}$ reproduces itself within the Dyson-Schwinger equation at the level of two skeleton diagrams for the polarization $\Pi$, and, thus, represents the proper effective paramagnon propagator. The linear in momentum term is generated by the second-order diagram, see $\Pi^{(1)}$ in the inset $(a)$ of Fig.\ref{Fig1}, while higher-order diagrams were argued to be small. This form of $\chi$ then leads to the FSD with flat hot spots because for renormalized spectrum
$\theta_N$ tends to zero. The major drawback of this scenario is an assumption that the renormalized coupling constant is small for which there is no evidence.

Using the diagMC technique for Eqs.~(\ref{HSF})-(\ref{SFI}) with $\chi^{-1}_0 = c(|\Delta {\bf q}_x| + |\Delta { \bf q}_y|) + \eta$, where $\eta \sim \xi^{-1}$ is the inverse correlation length, we computed the polarization and proper self-energy diagrams shown in both insets of Fig.~\ref{Fig1}, and confirmed the prediction of Ref.~\cite{Lee2017} regarding the formation of flat regions around hot spots in the limit of interest. However, the entire scheme turned out to be extremely fragile against finite temperature effects.

Fig.~\ref{Fig1} shows that already at $10^{-4} < T/t < 10^{-3}$ the momentum dependence of $\Pi^{(1)}$ deviates from the linear law. On the one hand this may explain why Ref.~\cite{Berg2020} did not observe it. On the other hand, our results invalidate the proposed in Ref.~\cite{Lee2017} FSD scenario at $T/t > 10^{-4}$. This ultra-low temperature scale finds no explanation in any of the microscopic system parameters or the effective theory of Ref.~\cite{Lee2017}. Thus, if FSD is observed in a microscopic model at higher temperature it has to be a different effective theory.

\begin{figure}[t]
\centerline{\includegraphics[scale=0.4]{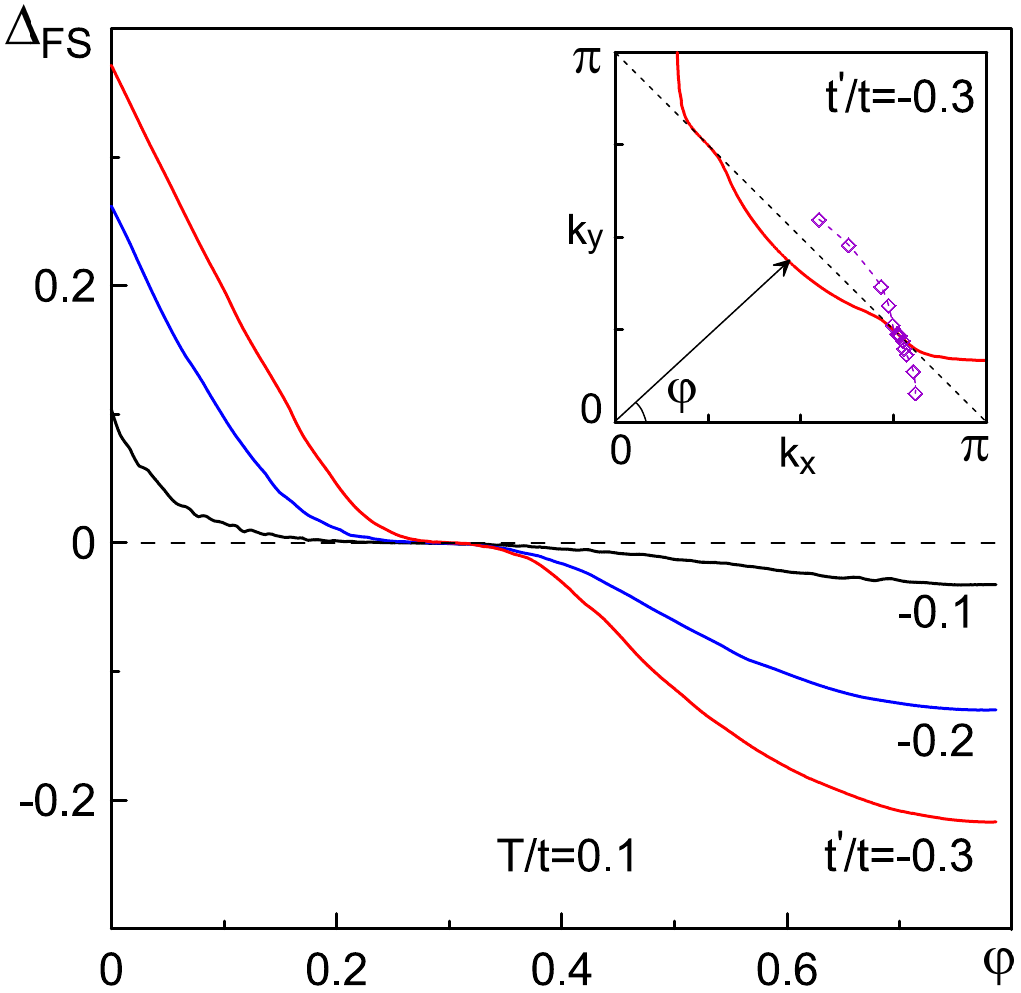}}
\caption{Deviation of the FS from the AFBZB (dashed line) in momentum space as a function of angle $\varphi$ for the SF model. The inset displays one segment of the FS (solid red line) extracted from $E_{\bf k}=0$ (see text). The dashed diamond curve shows part of the DLS corresponding to the maxima of $|\mathrm{Im} \Sigma^{(0)}_{\bf k}|$. All results are obtained for $T/t=0.1$, $c=1$, $g/t=0.4$, and $\eta = 10^{-3}$. }
\label{Fig2}
\end{figure}

If we proceed with computing the self-energy $\Sigma^{(0)}$ (see inset (b) in Fig.~\ref{Fig1}) with $\chi^{-1}_0 = c^2\Delta {\bf q}^2 + \eta^2$ and use it to determine the renormalized dispersion relation, $E_{\bf k}=\epsilon_{\bf k} + \mathrm{Re}\Sigma^{(0)}_{\bf k}(i\omega \to 0) - \mu$, counted from the chemical potential $\mu$, we obtain results for the FSD shown in Fig.~\ref{Fig2}. We plot the distance $\Delta_{FS}$ between the FS and AFBZB as a function of angle $\phi = \tan^{-1}[(k_y-\pi)/(k_x-\pi)]$ with $\Delta_{FS}=0$ corresponding to the hot spot. Clearly, it is possible to produce flat hot-spot regions with quadratic momentum dependence of $\chi$ at temperatures orders of magnitude higher than the limit established above for the linear momentum ansatz, including the $t'/t=-0.3$ case when the initial angle $\theta_N$ is relatively large.

When $\mathrm{Re}\Sigma_{\bf k}$ and $\mathrm{Im}\Sigma_{\bf k}$ have similar magnitudes, zeros of $E_{\bf k}$ may not coincide with spectral intensity peaks measured by ARPES. In this case, one may consider defining FS and the so-called Dzyaloshinskii-Luttinger surface (DLS) \cite{DLS2003,Fabrizio2022} from the maxima and minima of spectral intensity, respectively. The purple diamond curve in the inset of Fig.~\ref{Fig2} shows part of DLS extracted from the maxima of $|\mathrm{Im} \Sigma^{(0)}_{\bf k}|$ at the smallest Matsubara frequency, which also features a flat hot-spot. Within hot spots $|\mathrm{Im}\Sigma_{\bf k}|$ is large leading to suppressed intensity in the nested regions.

It was suggested in Ref.~\cite{Tsvelik-1-2017} that $\chi^{-1} \propto |\Delta {\bf q}|^2$ within the SF model confines quasiparticles within the flat FS regions, and, as a result, these parts of FS form a one dimensional subsystem weakly coupled to the rest of the FS. At sufficiently low temperature the state should be viewed as mixture of the non-Fermi liquid whose physics is dominated by strong one-dimensional fluctuations and the conventional Fermi liquid.
Eventually, the flat regions are gapped out leaving behind a state with reduced FS \cite{Tsvelik-2-2017,Tsvelik2019}.

Since the perturbative expansion for $\chi^{-1}_0 = c^2\Delta {\bf q}^2+\eta^2$ with vanishing $\eta$ breaks down, one has to assume that all vertex corrections are ultimately reduced to the renormalization of constants in the effective theory. The best way to verify this assumption and to establish a link between the microscopic physics and phenomenological treatments is to perform CDet simulations of the two-dimensional $(t-t')$ Hubbard model at half filling.

\smallskip

\noindent \textbf{Fermi-Hubbard model.} The Fermi-Hubbard model on the square lattice is the most studied prototypical model in the context of cuprate superconductors. Nevertheless, the possibility of the FSD towards flat hot spots in this model was never addressed by unbiased first-principles methods. Our approach is to employ the CDet technique at fixed density~\cite{SRF2022, simkovic2022origin} to evaluate Feynman diagrams up to $9$-th order to accurately compute the shape of the Fermi surface and spin susceptibility for the $(t-t')$ Hubbard model at half filling (for series resummation procedure see Supplemental Material \cite{SupplM}). The choice of half-filling was dictated by our goal of maximizing antiferromagnetic correlations and avoiding competing ordering tendencies.

\begin{figure}[t]
\centerline{\includegraphics[width=.48\textwidth]{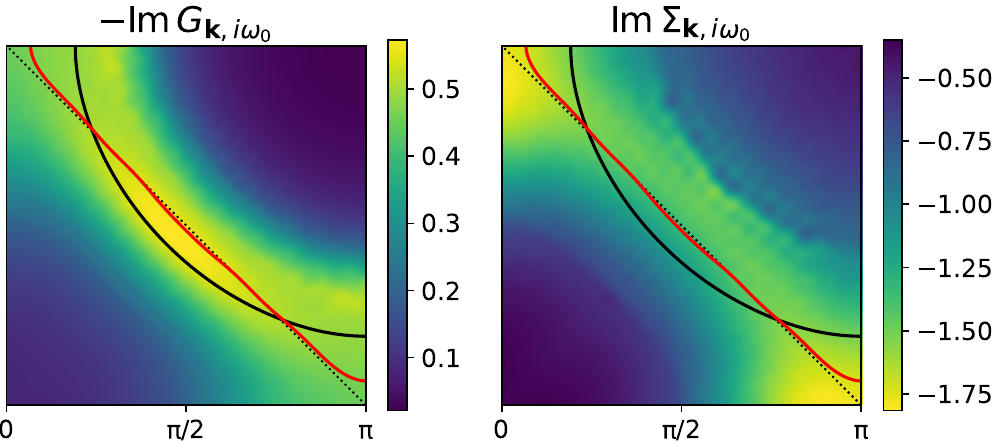}}
\caption{Momentum-resolved spectral function proxy (left) and imaginary part of the self-energy (right) at the lowest Matsubara frequency of the half-filled two-dimensional Hubbard model with parameters $U/t=5.75$, $t'/t=-0.3$, $T/t= 1/7$. The calculation was performed on a $60 \times 60$ lattice up to ninth order in the interaction strength, and, for visualization purposes, it was assumed that the self-energy is zero outside a $60 \times 60$ box around the origin. We plot here a quarter of the Brillouin zone. The black line is the non-interacting FS, while the red line is the solution of $E_{\mathbf k} =0$.}
\label{Fig3}
\end{figure}

We display in Fig.~\ref{Fig3} our results for a representative point $U/t=5.75$ and $T/t=1/7$. Defining the FS as the maximum of the spectral function proxy at the smallest Matsubara frequency, $-Im G_{\mathbf{k},i\omega_0}$, we see that it does not deform much. It is instead the line of zeros of $E_{\bf k}$ that undergoes significant deformation towards nesting. However, imaginary part of $\Sigma$ is large all along the AFBZB, and largest at the ``antinode'' $\mathbf{k}=(\pi,0)$ and the symmetry related points, as is clear from the second panel of Fig.~\ref{Fig3}. It suppresses the spectral weight along the AFBZB, indicating that the potential excitations suggested by the renormalized quasiparticle dispersion (which neglects $\mathrm{Im}\Sigma$) are in fact destroyed by large scattering with momentum transfer close to ${\bf Q}_N$. We emphasize that this is in strong contrast to what is expected from a weak-coupling antiferromagnetic spin-fluctuation theory, where one would expect that the maximum of the imaginary part of the self-energy would lie on a $(\pi,\pi)$-shifted version of FS. Hence, this nesting property of the self-energy is a strong-coupling non-perturbative effect. Correspondingly, it is close to the antinode that the real part of the self-energy has the strongest renormalization effect on the dispersion relation (see Supplemental Material \cite{SupplM}). We finally note that the effective field theory of Ref.~\cite{Lee2017} is not applicable at considered temperatures.

\begin{figure}[t]
\centerline{\subfigure{\includegraphics[scale=0.205]{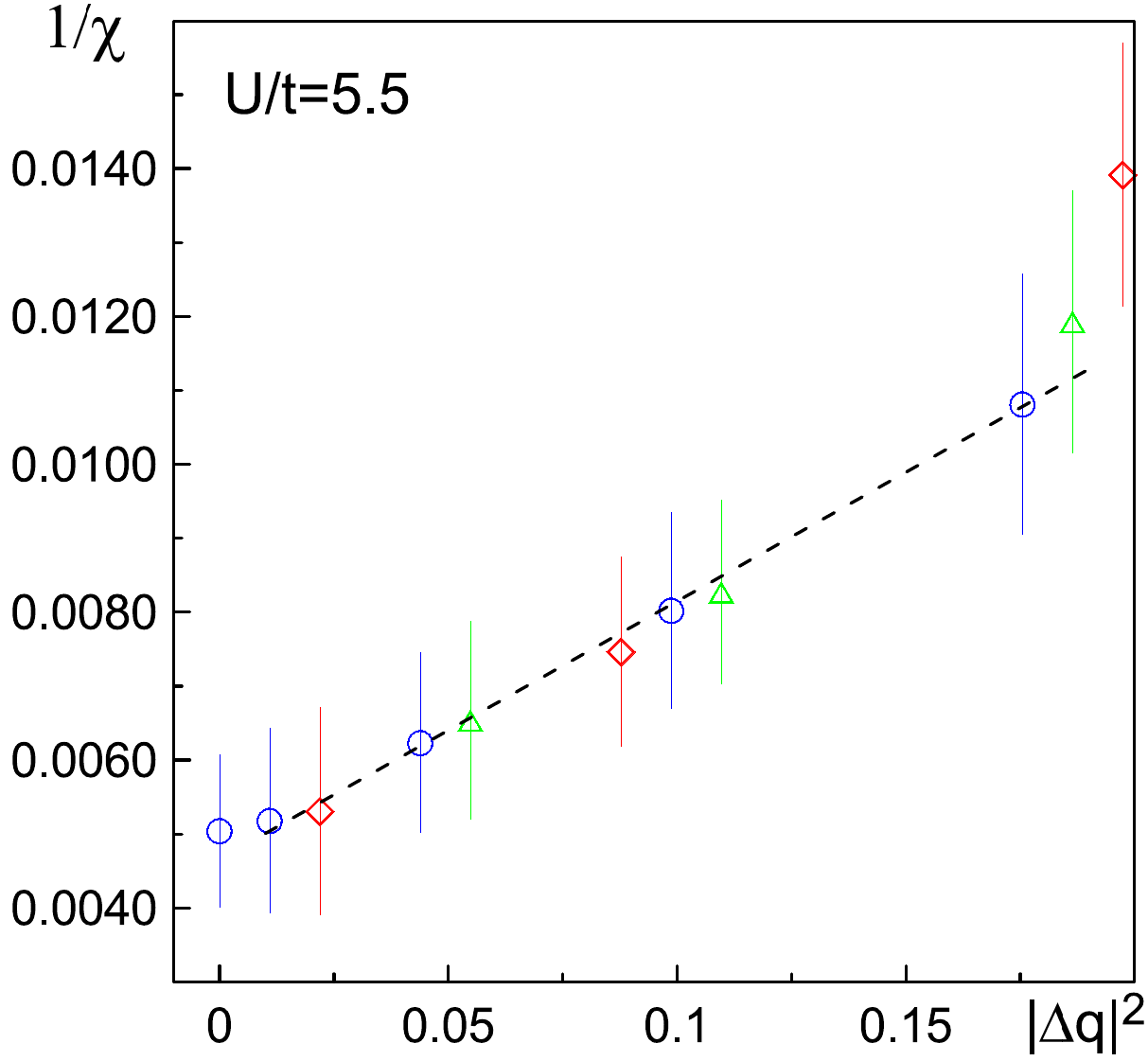}}
            \subfigure{\includegraphics[scale=0.205]{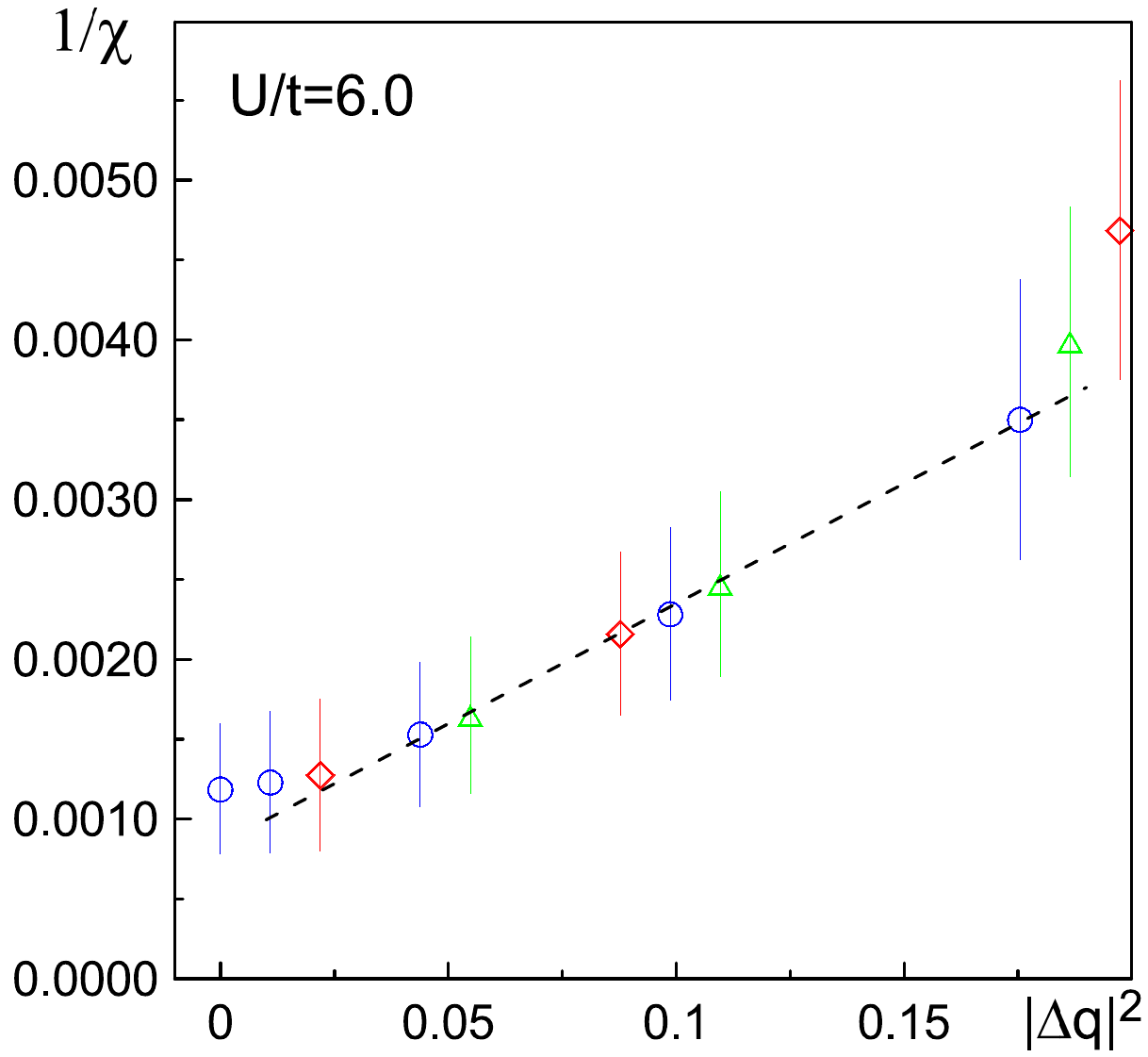}}}
\caption{Inverse static spin susceptibility as a function of $|\Delta {\bf q}|^2 \equiv |{\bf q}-{\bf Q}_N|^2$ for the $(t-t')$ Fermi-Hubbard model at half-filling and $U/t=5.5, 6.0$, $T/t=0.1$, $t'/t=-0.3$. Plots are presented for: (red diamonds) $(0,0) \to (\pi,\pi)$ direction; (blue circles) $(0,\pi) \to (\pi,\pi)$ direction; and (green triangles) an arbitrary direction to $(\pi,\pi)$. Dashed line is the fit demonstrating $a |\Delta q|^2 + b$ behavior of $\chi^{-1}$ at small deviations $|\Delta q|$ with momentum independent $a$ and $b$. (Reaching larger values of $|\Delta q|$ requires much longer simulation times.) }
\label{Fig4}
\end{figure}

The momentum dependence of the spin susceptibility is presented in Fig.~\ref{Fig4}. For small $|\Delta q|$ the response is nearly isotropic and closely follows the $c^2 |\Delta q|^2 + \xi^{-2}$ behavior with both $c$ and $\xi$ strongly dependent on $U$. At $U/t=6.5$ the correlation length is already about five lattice spacing. We take these results as direct confirmation that diverging spin correlations with quadratic dependence on $\Delta q$ are responsible for the formation of flat hot regions along AFBZB, in the $(t-t')$ Fermi-Hubbard model. Once such regions form, they effectively act as one-dimensional systems with non-Fermi liquid properties.


\smallskip


\noindent \textbf{Discussion and Conclusion.} Using diagrammatic Monte Carlo methods, we have considered different mechanisms and effective theories for formation of flat hot spots that amplify the nesting conditions in metals with near-critical anti-ferromagnetic correlations. For the SF model, we confirmed
predictions of the weak-coupling theory with linear in momentum dependence of the inverse paramagnon propagator $\chi^{-1}$. However, this linear dependence holds only at unreasonably low temperature. The quadratic momentum dependence of $\chi^{-1}$ is robust against finite-temperature effects and also leads to flat Fermi surface domains near the hot spots, but the corresponding effective field theory should assume that higher-order vertex corrections are already accounted for in the from of $\chi^{(0)}$.

High-order connected determinant Monte Carlo simulations of the $(t-t')$ Fermi-Hubbard model at half-filling reveal that the Fermi surface based on spectral intensity maxima does not undergo dramatic deformation towards flat AFBZB. It is instead the lines of zero of the renormalized quasiparticle dispersion that deforms towards nesting. This can be explained by a nesting property of the self energy, which becomes large along the AFBZB, a typical strong coupling effect. In the same parameter regime, $\chi$ has a sharp peak at the N\'{e}el wave vector ${\bf Q}_N = (\pm \pi, \pm \pi)$, and the best description of the dependence of $\chi^{-1}$ on momentum for small deviations from ${\bf Q}_N$ is quadratic.

The key common aspect between the SF and Hubbard model is that the quasiparticle dispersion is renormalized towards nesting. In the Hubbard model this, in turn, leads to enhanced scattering along the AFBZB which destroys quasiparticle coherence in this `hot' region; this effect being stronger near the antinodes (presumably due to the proximity of the van Hove singularity). In the spin-fermion model spectral intensity also gets suppressed  within flat hot spots.

At zero temperature, we expect from e.g. mean-field theory that the half-filled $(t-t')$ Hubbard model at intermediate values of $U$ has a metallic ground state with a FS reconstructed into hole and electron pockets due to antiferromagnetic ordering (at larger $U$, it becomes a Mott insulator with a fully gapped FS). Our results show that further understanding of how this state is reached as temperature is lowered and the correlation length grows, and of the corresponding fermiology at low but finite temperature, should take into account the quasi-one dimensional nature of nested hot regions, as was suggested in \cite{Tsvelik-1-2017}.

\smallskip

\noindent \textbf{Acknowledgements.} Authors are grateful to A. Chubukov for helpful discussions. I.S.T. and N.V.P. acknowledge support from the U.S. Department of Energy, Office of Science, Basic Energy Sciences, under Award DE-SC0023141. AMT was supported by Office of Basic Energy Sciences, Material Sciences and Engineering Division, U.S. Department of Energy (DOE) under Contracts No. DE-SC0012704. The Flatiron Institute is a division of the Simons Foundation.

\end{document}